\documentclass[twocolumn]{aastex6}

\shorttitle{The ESO SNPP Survey}
\shortauthors{Patat et al.}

\begin{document}


\title{The ESO Survey of Non-Publishing Programmes}


\author{F. Patat\altaffilmark{1}}
\author{H. M. J. Boffin}
\author{D. Bordelon}
\author{U. Grothkopf}
\author{S. Meakins}
\author{S. Mieske}
\author{M. Rejkuba}
\affil{European Southern Observatory\\
K. Schwarzschildstr. 2 \\
D-85748 Garching b. M\"unchen, Germany}


\altaffiltext{1}{fpatat@eso.org}

\begin{abstract}
One of the classic ways to measure the success of a scientific facility is the publication return, which is defined as the number of refereed papers produced per unit of allocated resources (for example, telescope time or proposals). The recent studies by \cite{sterzik15,sterzik16} have shown that 30-50 \% of the programmes allocated time at ESO do not produce a refereed publication. While this may be inherent to the scientific process, this finding prompted further investigation. For this purpose, ESO conducted a Survey of Non-Publishing Programmes (SNPP) within the activities of the Time Allocation Working Group\footnote{Time Allocation Working Group Report: \url{http://www.eso.org/public/about-eso/committees/ uc/uc-41st/TAWG_REPORT.pdf}}, similar to the monitoring campaign that was recently implemented at ALMA \citep{stoehr}. The SNPP targeted 1278 programmes scheduled between ESO Periods 78 and 90 (October 2006 to March 2013) that had not published a refereed paper as of April 2016. The poll was launched on 6 May 2016, remained open for four weeks, and returned 965 valid responses. This article summarises and discusses the results of this survey, the first of its kind at ESO.
\end{abstract}


\submitted{Draft version \today}



\section{Sample selection and general properties} \label{sec:intro}

The SNPP sample included all Normal, Guaranteed Time Observations (GTO) and Target of Opportunity (TOO) programmes that were scheduled between October 2006 and March 2013. This timeframe was selected to accommodate some delay between data acquisition and publication. To minimise ambiguity, we only considered programmes for which all runs were scheduled at the highest priority (i.e., Visitor Mode [VM] or A-ranked Service Mode [SM]). In addition, only programmes that had acquired a minimum amount of data were included in order to remove obvious cases, with a threshold of one science frame per allocated hour. In the selected period range, we identified 2716 proposals that obeyed the above criteria (90.7 \% of the total A-ranked SM and VM proposals), involving 2089 Normal, 478 GTO and 149 TOO programmes.
According to the ESO bibliographic database telbib\footnote{ESO telbib database: \url{http://telbib.eso.org}} \citep{uta}, 1278 (47.1 \%) of these programmes have not produced a refereed paper\footnote{Throughout this paper the definition of non-publishing programmes includes archival publications, i.e., articles that would be published by scientists not included in the list of co-investigators for the given proposal. Therefore, in this study, a non-publishing programme is one that has produced no refereed publication of any kind.} as of 16 April 2016. This gives an overall publication return of 52.9 \% with publication fractions of 52.5 \%, 52.7 \% and 59.7 \% for Normal, GTO and TOO programmes, respectively.

\begin{table*}
\tabcolsep 4mm
\centerline{
\begin{tabular}{lccccc}
\hline
\hline
Instrument & N. of Fractional 	& \% of total N. 	& N. of non-	                   & \% of total &	Non-Publishing \\
                  & Proposals & of Proposals                  & publishing Pr. &                   & Fraction (\%)\\
\hline
HARPS     &	103.4 &	3.8	& 22.9	& 1.8	& 22.1\\
FEROS     &	71.0	 & 2.6	& 21.8	&1.7	&30.7\\
MIDI      	   &  96.2	& 3.5 &	34.9	 & 2.7 &	36.3\\
UVES        &	148.6 &	5.5 &	56.0 &	4.4	& 37.7\\
EFOSC2    &	160.6 &	5.9 &	61.3	 & 4.8 &	38.2\\
SOFI      	&132.0 &	4.9 &	53.8	 & 4.2 &	40.8\\
FLAMES  &  	82.3	 & 3.0 &	34.9  &	2.7	& 42.5\\
FORS2     &	323.6 &	11.9 &	138.1 &	10.8 &	 42.7\\
FORS1   &  	55.7	 & 2.0 &	24.1	 & 1.9 &	43.3\\
EMMI      &	58.4	 & 2.2 &	25.7	 & 2.0 &	44.0\\
XSHOOTER & 	220.3 &	8.1 &	97.2 &	7.6 & 44.1\\
VISIR     & 	89.2	 & 3.3 &	43.4 & 	3.4	 &48.7\\
NACO      &	256.1 &	9.4 &	130.5	& 10.2 & 	51.0\\
OTHER     &	136.3 &	5.0 &	72.0	 & 5.6 &	52.8\\
HAWKI     &	51.7	 & 1.9 &	28.5	 & 2.2 &	55.1\\
SINFONI   &	123.5 &	4.5 &	68.9 & 	5.4 &	55.8\\
ISAAC     &	124.2 &	4.6	 & 71.9 & 	5.6 &	57.9\\
AMBER    & 	249.6 &	9.2	 &148.6	& 11.6 &	59.5\\
VIMOS     &	101.1 &	3.7 &	61.3	 & 4.8 &	60.7\\
CRIRES   & 	132.2 &	4.9 &	82.2	 & 6.4 &	62.1\\
\hline
ALL	& 2716.0	 & 100.0	 &1278.0 & 	100.0 &	47.1\\
\hline \hline
\end{tabular}
}
\caption{\label{tab:tab1}SNPP proposal distribution per instrument. The data are presented in ascending non-publishing fraction (last column). 
Only instruments with more than 50 programmes are listed separately. The rest is grouped under OTHER.}
\end{table*}

1143 Principal Investigators (PIs) were associated with the 2716 survey programmes; 755 (66.1 \%) of the PIs from this group did not publish a paper associated with these programmes. 34 \% of PIs published results for all programmes, 29 \% published results for some programmes, and 37 \% published results for none at all. 45 \% of the PIs were associated with only one programme from the survey, and 55 \% of these did not publish. On average, 1.1 proposals per PI have not yet produced a refereed paper.
The sample of 2716 survey programmes involves time allocated on 33 different instruments. For programmes that were allocated time on more than one instrument, we introduced the concept of a fractional proposal, attributing to a given instrument a fraction corresponding to the portion of total time assigned to it. For instance, if a programme was allocated one hour on FORS2 and four hours on UVES, this was counted as 0.2 and 0.8 proposals for the two instruments respectively. It is worth noting that 91.5 \% of the survey proposals requested time on a single instrument, and 7.6 \% requested two instruments. Table~\ref{tab:tab1} shows the distribution of proposals per instrument for the entire survey sample, as well as for the sub-sample that did not publish. For simplicity, we grouped instruments with fewer than 50 proposals under OTHER. These correspond to 5 \% of the total and involve eleven instruments, including SUSI2, TIMMI2 and VIRCAM.

Table~\ref{tab:tab1} also shows the nominal non- publishing fraction per instrument. According to this metric, which neglects instruments with low number statistics (i.e., OTHER), the most productive instrument is HARPS with a nominal publication return rate of about 78 \%. At the other end of the distribution, VIMOS and CRIRES are characterised by return rates lower than 39 \%. Although there is certainly a degree of instrument dependence, approximately 80 \% of the proposals show a publication rate of less than 60 \%, irrespective of the instrument used to produce the data.

\section{\label{sec:question}The questionnaire}

The PIs were asked the following question: "Why were you not able to publish the results of your observations in a refereed paper?" and were provided with ten possible options:

\begin{enumerate}
\item I did publish a refereed paper (provide a hyperlink in the comments).
\item Insufficient data quality (observations out of required specifications).
\item Insufficient data quantity (partially completed programme).
\item Inadequate ESO data reduction tools.
\item Null or inconclusive results.
\item Lack of resources on the PI side.
\item Science case no longer interesting.
\item I am still working on the data (provide time estimate in the comments).
\item I published a non-refereed paper (provide a hyperlink in the comments).
\item Other.
\end{enumerate}

\begin{figure}
\includegraphics[width=9cm]{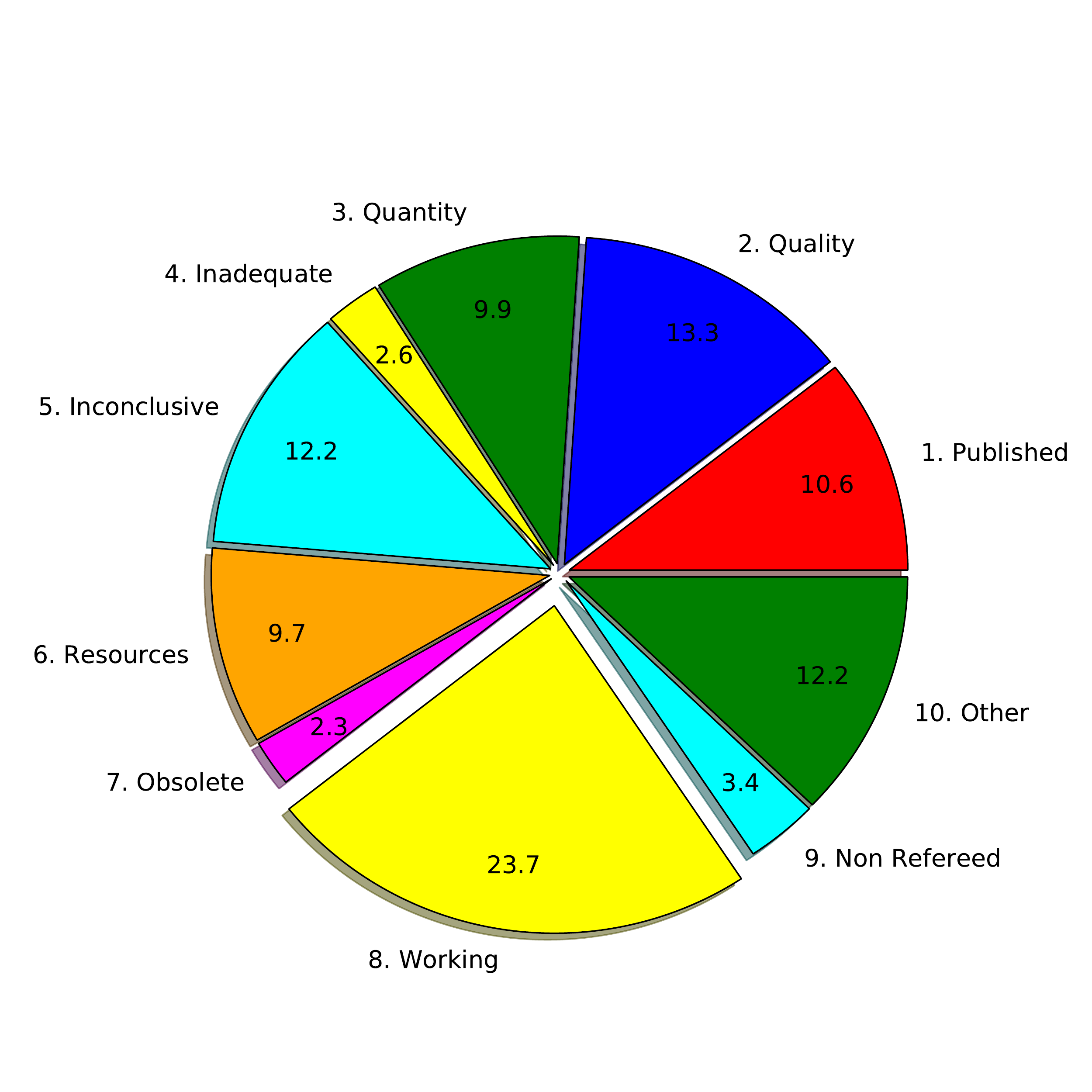}
\caption{\label{fig:fig1}Results of the SNPP survey (weighted fractions of the various options; see text).}
\end{figure}

The web form included a free-text field for comments. The responses were tagged with the Programme ID, to enable the analysis of correlations between the answer and programme properties (for example, time, constraints, instruments, scientific category, etc.).
Of the 1278 targeted programmes, we received responses for 965 (75.5 \%). Accounting for the fact that approximately 70 queries could not be delivered (due
to out-of-date User Portal profiles), the response return was 80 \%, which is much higher than expected from web-based surveys ($\sim$10 \%; \citet{fan}). The response rate increased for more recent time allocations, with a response rate of 85 \% from PIs associated with programmes from the last semester, compared to 70 \% from PIs from the first semester. PIs were allowed to select more than one option in their replies. Most selected a single option (55.5 \%), with 31.1 \% selecting two options and fewer than 10 \% selecting three. The most popular single-option response was "8. I am still working on the data" (14 \%), followed by "1. I did publish a refereed paper" (9 \%). The most popular two-option response was "6. Lack of resources on the PI side" and "8. I am still working on the data" (5 \%), followed by "2. Insufficient data quality" and "3. Insufficient data quantity" (3 \%). The general outcome of the survey is summarised in Table~\ref{tab:tab2}. Given the possible multiple options within each single response, the results are presented in two flavours. For each single option, we list the number and percentage of responses and the weighted number and percentage. The weighted values were computed by giving equal weights to the various options within the same response (Figure~\ref{fig:fig1}). By construction, the number of weighted responses (and percentages) adds up to 965 (100 \%), whereas this is obviously not the case for the non-weighted responses. The two sets of numbers have different meanings: the latter is related to the frequency of responses associated with a given option, while the former provides information about its relative importance. The difference becomes clearer when considering the following simplified example. If a hypothetical survey includes the following four responses: (1, 1, [2, 4, 5, 6], [2, 7, 8, 9]), the non-weighted frequencies of options 1 and 2 are both 50 \%. On the other hand, the weighted fractions of the two options are 50 \% (1) and 25 \% (2), respectively. Therefore, while options 1 and 2 were included in the same fraction of responses (50 \%), option 1 is twice as significant.
The breakdown of responses by instrument shows some instrument-specific dependencies. For instance, while for X-shooter the frequency of option 8 is equal to the average (23.7 \%), UVES is characterised by a significantly larger fraction (35.5 \%), and AMBER shows a lower fraction (18.0 \%). This may be related to the specific scientific areas covered by the instruments, the complexity of the science cases involved, and their appeal to the community.
In the following sections, we will go into more detail for each of the options in the questionnaire.

\begin{table*}
\centerline{
\tabcolsep 5mm
\begin{tabular}{lcccc}
\hline \hline
               & \multicolumn{2}{c}{Reponses including} & \multicolumn{2}{c}{Weighted responses}\\
               &\multicolumn{2}{c}{options} & & \\
Option	&	No.	&	\%	&	No.	&	\%	\\
\hline
1. Published	&	124	&	12.8	&	102.5	&	10.6	\\
2. Insufficient quality	&	202	&	20.9	&	128.7	&	13.3	\\
3. Insufficient quantity	&	165	&	17.1	&	95.3	&	9.9	\\
4. Inadequate tools	&	61	&	6.3	&	25.2	&	2.6	\\
5. Null or inconclusive	&	187	&	19.4	&	117.9	&	12.2	\\
6. Lack for resources	&	176	&	18.2	&	93.8	&	9.7	\\
7. No longer interesting	&	38	&	3.9	&	21.9	&	2.3	\\
8. Still working	&	352	&	36.5	&	228.3	&	23.7	\\
9. Non-refereed paper	&	66	&	6.8	&	33.2	&	3.4	\\
10. Other	&	188	&	19.5	&	118.2	&	12.2	\\
\hline
	&		&		&	965	&	100.0	\\
\hline \hline
\end{tabular}
}
\caption{\label{tab:tab2}Summary of the SNPP responses.}
\end{table*}

\subsection{Option 1: I did publish a refereed paper}

Of the 124 responses associated with option 1, 14 provided incomplete information (for example, no link to the refereed publication or a link to a non-refereed publication). These cases were conservatively counted as non-published. The remaining 110 replies can be grouped as follows: a) the Programme ID was either wrong or absent (61; 55.5 \%); b) the refereed paper appeared in print after the SNPP sample definition and was listed by telbib (25; 22.7 \%); c) the paper is in the process of being accepted (21; 19.1 \%); and d) the paper is missing from telbib (3; 2.7 \%). 11.4 \% of the responses correspond to false negatives (i.e., published programmes that were initially classified as non-publishing). This fraction, deduced from the 965 replies, can be used to compute the completeness-corrected value of the publication rate within the whole SNPP sample (N = 2716; 58.9 \%).
In the following we use the term "completeness" to refer to the completeness of telbib. In response to the information provided by the PIs, 64 telbib records were modified. The vast majority (87.5 \%) of these records were previously included in the database, but the particular Programme ID in question was missing. We updated these records accordingly. Only eight papers (12.5 \%) had not previously been considered as using ESO data; these records were added to the telbib database without further verification.
As a side note, the SNPP has allowed us to robustly determine that the telbib completeness is better than 96 \%.

\subsection{Options 2 and 3: Insufficient data quality and/or quantity}

We will discuss options 2 and 3 together because there is a clear overlap, as confirmed by comments from the PIs. In total, these two options account for 23.2 \% of the cases, with 8.2 \% citing only option 2, and 4.9 \% citing only option 3. There is a striking difference between SM (32 \%) and VM (68 \%) programmes in the responses associated with option 2. This is likely due to the fact that VM observations are more adversely affected by bad weather conditions, while by definition SM is less affected by weather. We found a small correlation with requested seeing constraint and Quality Control (QC) grades in the SM programmes. Unsurprisingly, the majority of the affected SM programmes requested relatively good conditions (seeing < 1 arcsecond) and associated observations had higher fractions of B quality control (QC) grades (i.e., one of the observing constraints was violated up to 10 \%) compared to the rest of the sample.
A clear dichotomy is also seen when considering responses per telescope, with the largest fractions related to the Very Large Telescope Interferometer (VLTI; 26 \%), and Unit Telescope 1 (UT1; 20 \%). The vast majority (90 \%) of VLTI programmes involved AMBER and were associated with Guaranteed Time Observations (GTO), which are often riskier as they tend to involve new instrumentation. For UT1, most cases are related to the early years of CRIRES operations or problems with the degraded coating of the FORS2 longitudinal atmospheric dispersion corrector, which have since been resolved \citep{boffin}.
A detailed analysis of the responses that only cited option 3 confirms that the corresponding programmes had been affected by weather, technical losses (in VM), or a completion fraction of lower than $\sim$50 \% (for SM). We conclude that most of the cases involving options 2 and 3 can be accounted for within ESO's operation model, and/or reflect the early operation of new complex systems.

\subsection{Option 4: Inadequate ESO tools}

This was the least selected option, with a weighted fraction below 3 \%, indicating that a negligible fraction of users identify the software provided by ESO as the cause for non-publication.

\subsection{Option 5: Null or inconclusive results}

The fraction of cases reporting null or inconclusive results is comparable to that of option 2 (insufficient quality). Although null or inconclusive results are arguably part of the scientific process, PIs may be reluctant to admit this, potentially biasing the responses and underestimating
the fraction. No correlation was found between the fraction of inconclusive results and the scientific subcategories
of the programmes, indicating that all science cases are affected in similar ways.

\subsection{Option 6: Lack of PI resources}

The weighted frequency of this option is 9.7 \%. When considered together with option 8 below, these two options account for 33.4 \% and point to a significant difficulty in the community to keep up with the rate of data production.

\subsection{Option 7: Science case no longer interesting}
    
Only 2.3 \% of the cases were indicated as obsolete science. These occurrences can be tentatively identified as instances in which the data delivery duty cycle and/or the time taken for the PI to make the data publishable was too long compared to the evolution in the given field.   

\begin{figure}
\includegraphics[width=9cm]{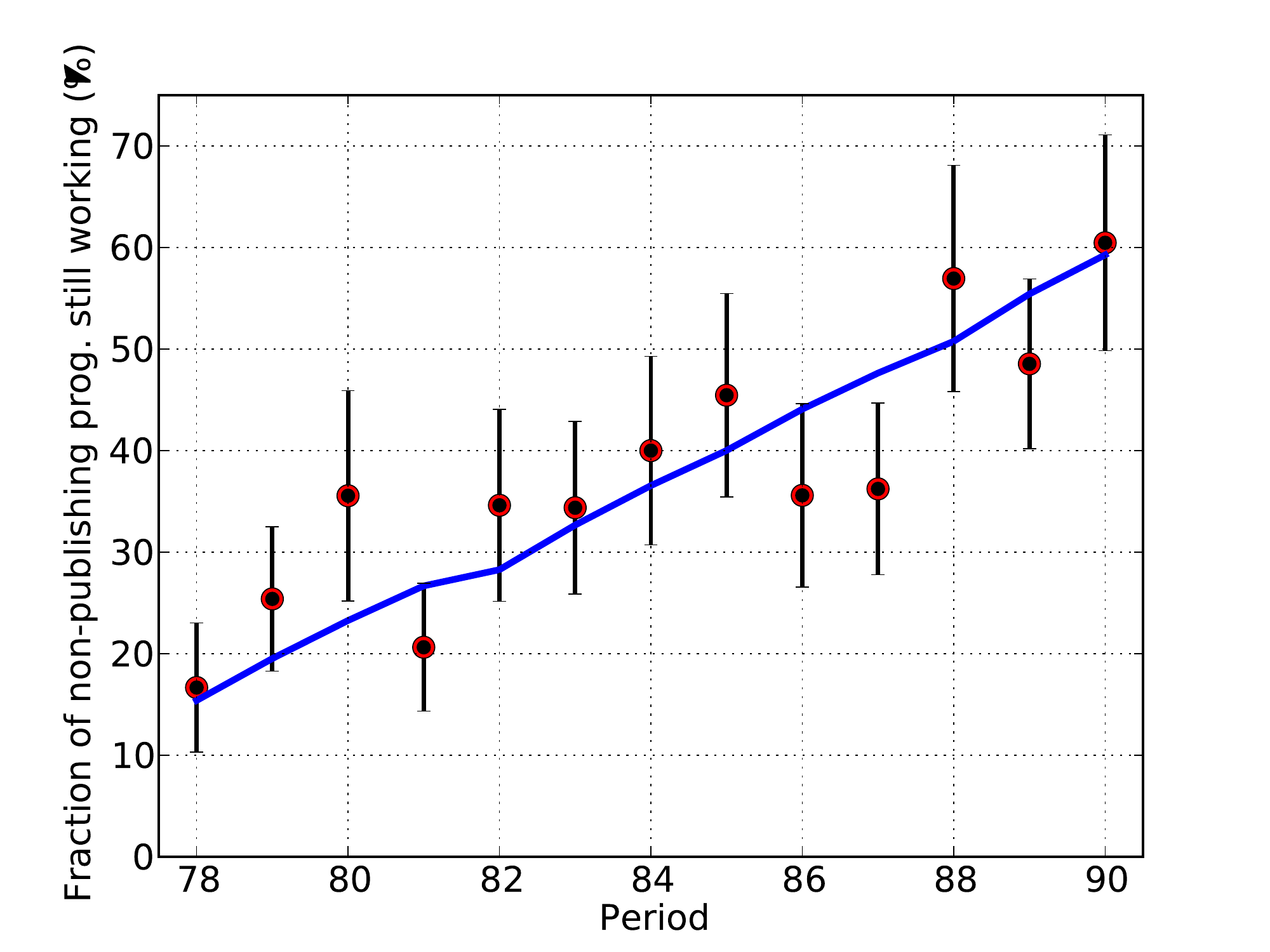}
\caption{\label{fig:fig2}Completeness-corrected fraction of non-publishing programmes still working on the data as a function of allocation period (red symbols). The blue line traces the expected behaviour for $f^0_P$ = 75.2 \% (see text). The error bars indicate the Poissonian uncertainties (1-sigma level).}
\end{figure}

\subsection{Option 8: I am still working on the data} 

This was the most frequent response. Excluding the 13 cases in which options 1 and 8 were selected, a total of 339 responses included this option: 135 as single option, 49 with option 6, 26 with option 10, and 129 in other combinations. For a more quantitative approach we introduce the ratio, R, between the number of proposals for which work is still in progress and the total number of non-publishing proposals (corrected for telbib completeness). The previous numbers yield R = 339/(965-110) = 39.6$\pm$2.5 \%
for the overall SNPP sample. This ratio can be calculated individually for each semester to study its evolution with time. The completeness-corrected result is presented in Figure~\ref{fig:fig2}, which shows a net and steady overall decrease for older programmes. The fact that R = 78 and not zero for the earliest semester in the sample indicates that it takes longer than 12 semesters for all programmes that will eventually produce a refereed publication to do so.

\begin{figure}
\centerline{
\includegraphics[width=9cm]{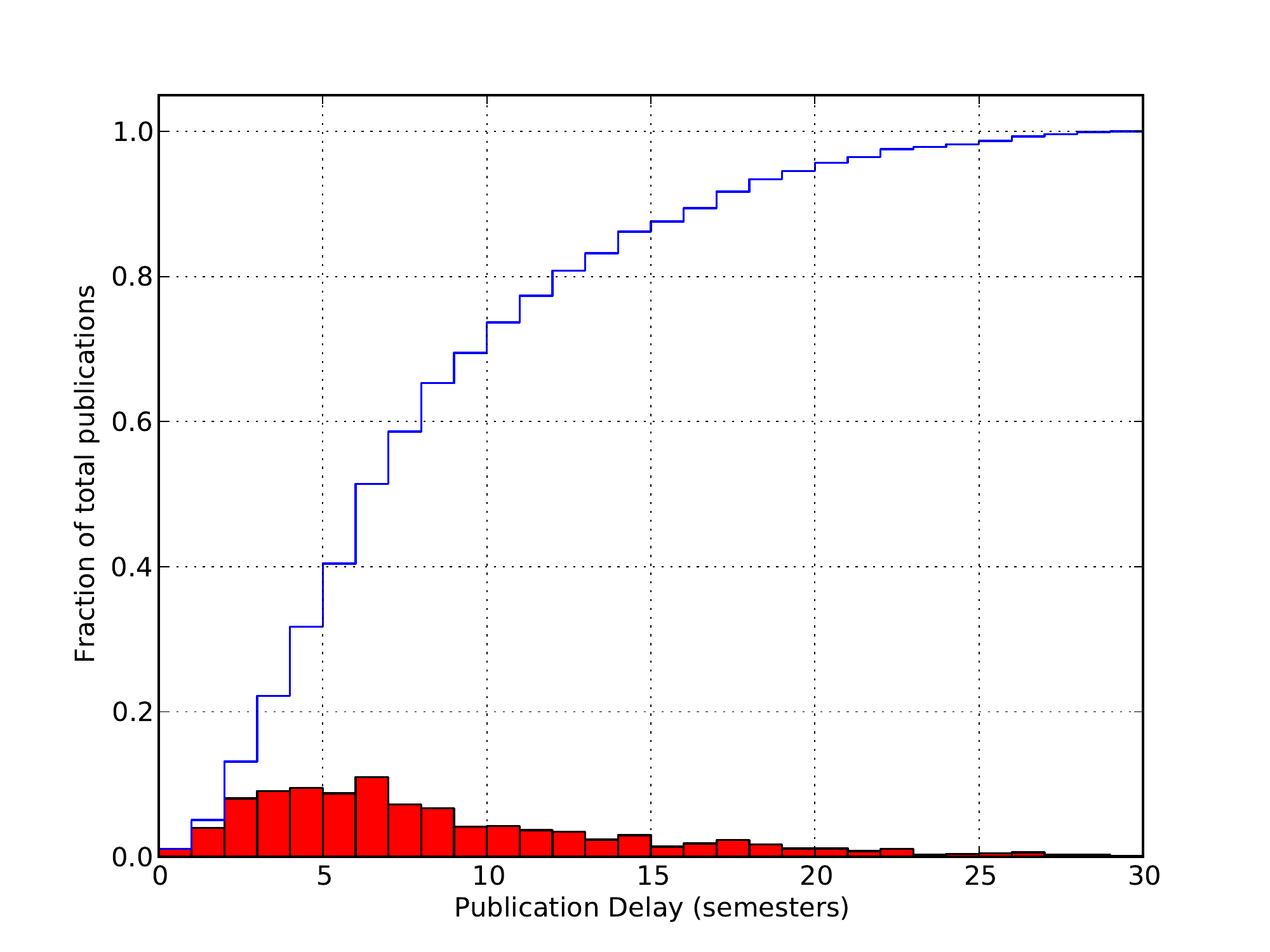}
}
\caption{\label{fig:fig3}Publication delay time distribution. The blue line traces the cumulative distribution function $C(t)$.}
\end{figure}

Before we discuss this result in more detail, we will define the Publication Delay Time Distribution (PDTD), which describes the delay between the allocation and the publication time. This provides a measure of the complete duty cycle, including the time for ESO to deliver the data and for the user to process, analyse and publish them.
We used the data provided by the ESO telbib interface to derive this function. For each year from 2008-2015 we extracted the refereed publications per programme for programmes that used Paranal telescopes. Due to their nature, Large Programmes and Director Discretionary Time Proposals were excluded. Each publication in the sample of 1303 refereed papers is characterised by the publication year ($t_P$) and the programme's allocation period (P). A given publication year is tagged with its central semester\footnote{Any given calendar year intersects with three ESO semesters, only one of which is fully contained in the given year (the one running from 1 April to 1 October). We call this the central semester P0. The data show that for any year in our telbib sample, no publication is produced in $P_0+1$, while there is always at least one publication in $P_0$, and several in $P_0?1$. Therefore, $P_0$ can be regarded as the most recent scheduled period producing a publication in the given year. $P_0$ is simply given by
$P_0 = 2(t_P-2008)+81$. Although the time delay could be defined and computed in a more accurate way, a resolution of one semester is sufficient for the purposes of this study.}, $P_0$. The publication delay, in semesters, is then computed as $\Delta P = P-P_0$.
The sample data show that only 1.1 \% of papers are published with a null delay using the above definition, while this grows to 11 \% for $\Delta P$=6 semesters, after which it steadily decreases for larger delays. This is illustrated in Figure~\ref{fig:fig3}, which also includes the cumulative distribution function, indicated as $C(t)$ (where the time $t $is counted from $P$). At face value, it takes 7 semesters to reach 50 \% of the publications, and 20 semesters
to reach 95 \%, in agreement with \cite{sterzik16}. The quantity $1-C(t)$ can be regarded as the probability that a programme that has not published a refereed paper after a time $t$, will publish it in the future. For example, a programme that has not published after 10 semesters has a 22 \% residual probability of publishing in the future.

\begin{table*}
\tabcolsep 3mm
\centerline{
\begin{tabular}{c|cc|c|cc|c}
\hline
              & \multicolumn{3}{c}{Service Mode} & \multicolumn{3}{c}{Visitor Mode} \\ \cline{2-7}
              & \multicolumn{2}{c}{Allocated time} & Publishing & \multicolumn{2}{c}{Allocated time} & Publishing\\ \cline{2-3} \cline{5-6}
Quartile & Time range & Median time & fraction (\%) & Time range & Media time &  fraction (\%)\\
              & (nights)       & (nights)         &  & (nights)    & (nights)   & \\
\hline
1 &	0.1-0.8	& 0.4 &	39.5$\pm$4.3 &	0.1-1.2	& 1.0	& 50.5$\pm$4.5\\
2 &	0.8-1.4	& 1.0 &	53.6$\pm$4.7 &	1.2-2.1	 & 2.0	& 53.0$\pm$4.7\\
3 &	1.4-2.4	& 1.9 &	58.9$\pm$5.9 &	2.1-4.0	 & 3.0	& 66.0$\pm$4.8\\
4 &	2.4-12.5	& 3.4 &	61.3$\pm$6.1 &	4.0-12.5	 & 6.0	& 68.9$\pm$6.5\\
\hline \hline
\end{tabular}
}
\caption{\label{tab:tab3}Fraction of proposals that published at least one refereed paper for Service and Visitor Mode programmes as a function 
of allocated time (in nights) in the four quartiles of the respective time distributions.}
\end{table*}

The behaviour of $R(t)$ in Figure~\ref{fig:fig2} is a direct consequence of the publication delay. In fact, it is easy to show (see Appendix~\ref{sec:app}) that if $f^0_P$ is the underlying publication fraction (i.e., the return rate one would measure for a sample of programmes at a time when $C=1$), then the ratio $R(t)$ observed for a set of proposals all allocated in the same period and observed after a time t (i.e., the time when the survey is carried out) is given by:

\begin{displaymath}
R(t) = f^0_P \; \frac{1-C(t)}{1-f^0_P\; C(t)}
\end{displaymath}

One can also show (see Appendix~\ref{sec:app}) that this expression can be applied to compute $\bar{R}$ for a whole sample, including programmes allocated in a period range, by replacing $C(t)$ with its weighted average $\bar{C}$:

\begin{displaymath}
\bar{C} = \frac{\sum_P  N(P) \; C(P_S-P)}{\sum_P N(P)}
\end{displaymath}

where $N(P)$ is the number of proposals allocated in semester $P$, and $P_S$ is the period in which the survey is run. It can be readily demonstrated (see Appendix~\ref{sec:app}) that the expected publication fraction at the time of the survey is simply $f_P= f^0_P \; \bar{C}$. In the real case $C = 0.78$, while the SNPP
provided $\bar{R}$ = 0.396$\pm$0.025. The above relation can be inverted to express $f^0_P$ as a function of $\bar{R}$, from which one can finally estimate the delay- and completeness-corrected return rate: $f^0_P$ = 0.75$\pm$0.01. This implies that after waiting a sufficiently long time, more than 20 semesters after the most recent period in the sample (see Figure~\ref{fig:fig3}), one would measure a publication return of approximately 75 \%.
This calculation conservatively assumes that all programmes for which the users have specified option 8 will eventually publish. This assumption can be verified by comparing the real data with two predictions that descend from the above equation. The first is the overall publication fraction expected for the real SNPP case, which is given by $f_P = f^0_P \; \bar{C}$ = 58.5$\pm$1.0 \%. This can be directly compared to the completeness-corrected value derived from SNPP, 58.9 \% (see above), which is fully consistent within the estimated uncertainty. 

The second prediction concerns the time dependence of $R(t)$, as defined by the above relation. This is compared to the real SNPP data in Figure~\ref{fig:fig2} (blue line), which illustrates how the predicted behaviour matches the data within the estimated uncertainties.
The above results indicate that the SNPP fraction of option 8 gives a realistic representation of the situation and is not the result of a "convenient answer" from PIs attempting to justify a lack of publication. In other words, the SNPP result is fully compatible with the estimated PDTD, and shows that the publication delay correction is significant, especially when the most recent periods included in the sample date back less than 10-12 semesters at the time of the survey.

\subsection{Option 9: I published a non-refereed paper}

The cases in which a programme did not publish a refereed paper but rather a non-refereed article account only for 3.5 \% of the total. This implies that, with very few exceptions, if a project does not produce a refereed publication then it will not produce any publication at all.

\subsection{Option 10: Other}

This option reflected 12.3 \% of the cases and the associated comments yielded a mixture of reasons, the most frequent being that the person leading the project left the field. Other recurrent explanations included: lack of ancillary data from other facilities, results not meeting expectations, lowered priority of the project because of more pressing activities, quicker results obtained by other teams and/or with better-suited instruments, non-detections, etc.

\section{Considerations on observing mode and allocated time}

\begin{figure}
\centerline{
\includegraphics[width=9cm]{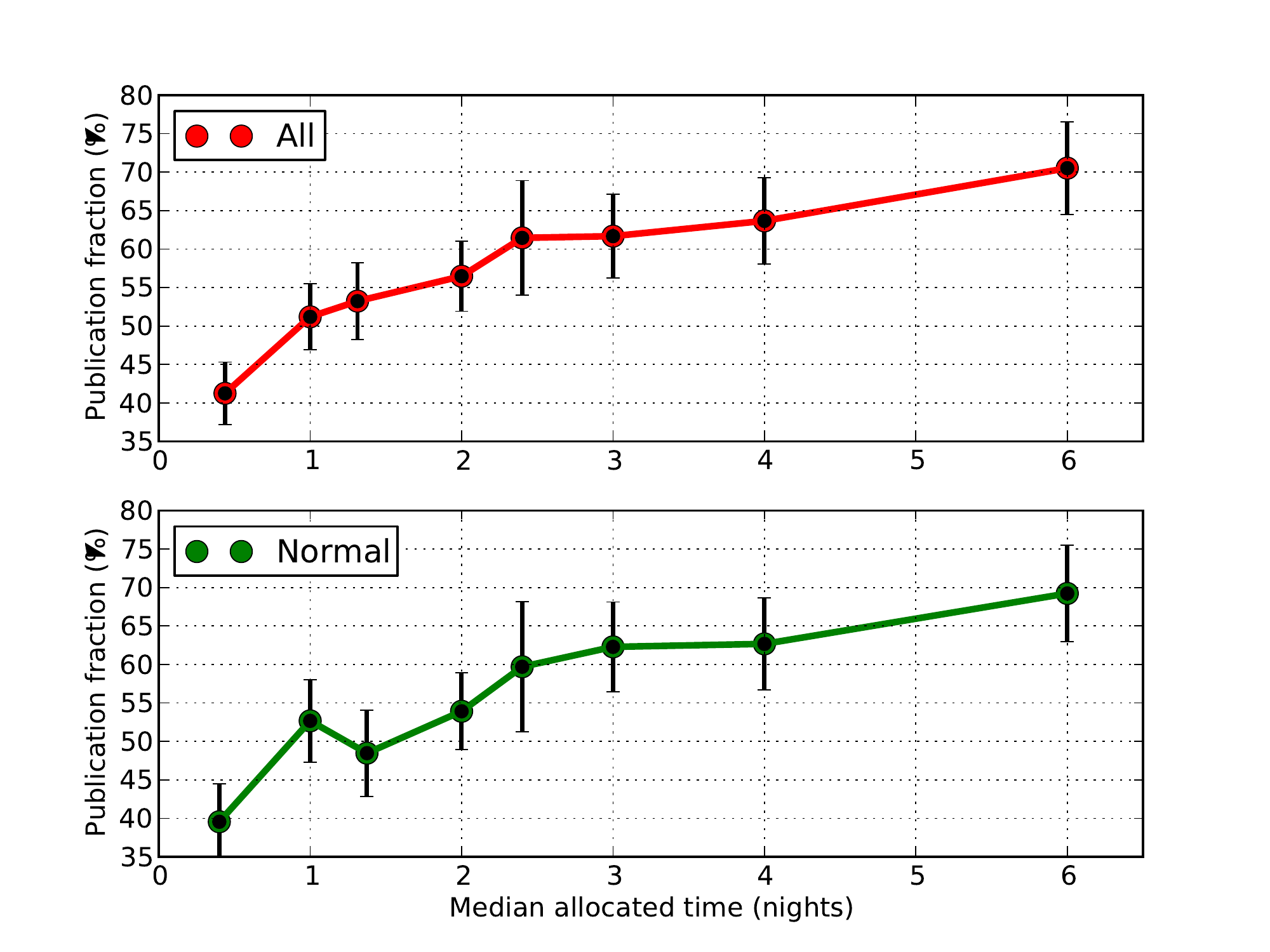}
}
\caption{\label{fig:fig4}Fraction of proposals that publish at least one refereed paper as a function of allocated time (nights) for all programme types (upper panel) and for normal programmes only (lower panel). The ×-axis position of the points marks the median allocated time within each octile bin.}
\end{figure}

As a final analysis, we have derived the completeness-corrected publication fractions considering VM and SM separately, as the two observing modes were reported to behave in a different way by \cite{sterzik16}. For this purpose, we have considered only single observing mode proposals within the SNPP initial sample, including 1089 SM programmes (40.1 \%) and 1493 VM programmes (55.0 \%). The remaining 134 mixed observing mode programmes (4.9 \%) were excluded from the calculations. For each of the observing modes we have computed the time intervals that define the four quartiles of the respective time distributions. These differ for SM and VM, with median allocated times of 1.4 and 2.1 nights, respectively. For the time conversion, we adopted the ESO convention of 10 hours per night in odd periods and 8 hours per night in even periods.
Finally, we derived the publication fraction, $f_P$, within each time bin for the two observing modes separately (see Table~\ref{tab:tab3}). An interesting feature, common to both SM and VM, is the steady increase of the return rate for larger time allocations: the publication fractions in the fourth quartile are 60 \% and 40 \% larger than in the first quartile for the two modes, respectively.
Another aspect is the larger return of VM programmes when compared to SM \citep{sterzik16}. To some extent this is expected, as VM programmes tend to be larger than SM programmes. This becomes clearer when comparing SM and VM runs with the same median duration. For instance, the two rates are very similar for SM runs in their second quartile (53.6 \%) and the VM runs in their first quartile (50.5 \%), both having a median duration of one night. Although observing mode effects cannot be excluded, the amount of time allocated to the programme appears to be the dominant factor. Figure~\ref{fig:fig4} (upper panel) shows the dependence of publications on the allocated time, plotting the completeness-corrected publication fraction measured by SNPP in octiles of the overall time distribution (each time bin includes about 320 proposals). GTO programmes constitute 17.6 \% of this sample, potentially biasing this result. As GTO programmes make systematic use of novel instruments designed to cover the specific science cases for which they were built, they tend to be more productive than average \citep{sterzik16}. For this reason, we produced a similar plot for Normal programmes (Figure~\ref{fig:fig4}, lower panel), which reveals a similar trend, albeit with more noise. We conclude that larger programmes tend to be more productive on average; this is in line with the results of \citet{sterzik15, sterzik16}. We find the same trend within the Normal programmes, which account for the largest fraction of the allocation (both in terms of number of proposals and time).
In an attempt to understand what makes larger allotments more productive, we examined the frequency of the SNPP options as a function of allocated time, dividing the programmes into the four quartiles of the time distribution. No significant dependence was found for any of the options, suggesting that the lower observed return rate $f_P$ for smaller time allocations was the fruit of a lower inherent return rate $f^0_P$, regardless of the reason for the lack of publication.
We note that two non-publishing programmes with very different allocations are counted in the same way here. However, it is obvious that they have a different impact in terms of "wasted" telescope time. To quantify this aspect, we computed the telbib completeness-corrected fraction of scheduled time that was allocated to non-publishing programmes as a function of their size (in the four quartiles of the time distribution). We did this for the entire SNPP sample, both as observed and correcting for the publication delay (Table~\ref{tab:tab4}), assuming that all the work from in progress cases will eventually produce a refereed paper. At the time of the SNPP survey, about 37 \% of the time allocated to A-ranked SM and VM programmes had not produced a refereed publication. This fraction in time is very similar to the corresponding completeness-corrected fraction in proposals (100 \% - 58.9 \% = 41 \%). Once corrected for the publication delay, this fraction reduces to about 25 \%. Therefore, as in the case of the number of proposals, about one quarter of the telescope time allotted to A-ranked SM and VM proposals will not lead to a refereed publication.

A closer inspection of Table~\ref{tab:tab4} reveals that, although larger programmes tend to be more successful in terms of producing at least one publication (Table~\ref{tab:tab3} and Figure~\ref{fig:fig4}), the non-publishing time fraction tends to increase with their size. This finding is equivalent to the lower number of publications per programme per unit of allocated time that was reported by \cite{sterzik15, sterzik16} for proposals with sizes between the short Normal (below 20 hours) and Large Programmes (above 100 hours).

\begin{table*}
\tabcolsep 8mm
\centerline{
\begin{tabular}{ccccc}
\hline \hline
              &  \multicolumn{2}{c}{Allocated time}  &  \multicolumn{2}{c}{Fraction of total time allocated} \\ \cline{2-3}
Quartile & Time range & Median time & \multicolumn{2}{c}{to non-publishing progs. (\%)} \\ \cline{4-5}
              & (nights) & (nights) & Observed & Delay-corrected \\
\hline
1 &	0.1-1.0 &	0.7	& 4.0 &	3.1\\
2 &	1.0-2.0 &	1.7	& 8.5 &	6.1\\
3 &	2.0-3.0 &	3.0	 & 8.1 &	5.6\\
4 &	3.0-12.5	& 4.5	& 16.2 &	10.5\\
\hline
All &	0.1-12.5 &	2.0 &	36.8 & 	25.3\\
\hline \hline
\end{tabular}
}
\caption{\label{tab:tab4}Fraction of allocated telescope time not producing a refereed paper in the four quartiles of the time distribution measured by SNPP (Observed) and extrapolated in the hypothesis that all programmes that included option 8 (still working) will eventually publish (Delay-corrected).}
\end{table*}

One can assume that there exists an optimal distribution of allocated times that maximises scientific return and minimises the waste of telescope time. Identifying such an ideal distribution is beyond the scope of this paper. However, Table~\ref{tab:tab4} allows us to gain a first insight into the boundary conditions of such a parameter search: in both cases (observed and delay-corrected), programmes with allocations below and above $\sim$2.5 nights "waste" the same amount of time. This implies that increasing the number of programmes with allocations larger than this value would effectively decrease the overall amount of time that leads to no refereed publication.
This can be understood considering two extreme cases in which the schedule is completely filled with a) only programmes shorter than one night, or b) only programmes longer than three nights. The first case would yield a much larger number of allocated programmes than in the second case (by a factor larger than 12), but the total amount of "wasted" time would also be larger. The SNPP data, once corrected for completeness and time delay, show that about 40 \% of programmes shorter than one night do not publish, producing a time waste of this same magnitude in the hypothetical first case.
On the other hand, programmes longer than three nights would "waste" less time (about 20 \%), but the number of published papers would be much smaller than in the first case, which would likely result in a decrease of the overall scientific return. These simple considerations suggest that the optimal distribution of allocated times must ensure the proper level of diversity, by including a mix of programme sizes.

\section{Conclusions}

The performance of a scientific facility can be evaluated using various metrics, each of which are affected by different issues. In this study, we have focused on the binary bibliographic figure of merit, i.e., the publication or lack of publication of at least one refereed paper. This is one of the simplest bibliometric estimators, as it does not account for the publication's impact or the resources involved. The fact that a programme has not yielded a refereed publication does not necessarily imply that the observations were a complete waste of resources. Nevertheless, analysing this aspect and understanding its possible causes is certainly one of the basic steps that institutes and organisations such as ESO must undertake to characterise their overall efficiency.
The SNPP has shown that there are many reasons why a programme may not produce a refereed publication. With the notable exception of option 8 ("team still working on the data") and the combination of options 2 and 3 ("insufficient data quality and quantity"), there is not a single, dominant culprit.
The relatively large fraction of proposals for which work is still in progress ($\sim$40 \%) is fully compatible with the Publication Delay Time Distribution deduced from an independent set of programmes. Once corrected for the publication completeness of the telbib database - where the vast majority of the missing cases are generated by wrong or absent Programme IDs in the published papers - and for the publication delay, the estimated asymptotic publication rate is approximately 75 \%. This means that, at least in the phase covered by the SNPP, about a quarter of the proposals scheduled in VM and/or in A-ranked SM will never publish a refereed paper.
Although this fraction can likely be decreased by further improving the overall workflow, part of the problem may be inherent. The non-negligible fraction of cases of insufficient resources (generally option 6 but also indicated in option 10) and the typically long publication delay may be symptoms of workload pressure in the community. The significant numbers of cases in which negative or inconclusive results do not turn into publications also support this conclusion. This reflects what may be a growing cultural problem in the community as scientists tend to concentrate on appealing results, especially if they have limited resources, and need to focus predominantly on projects that promise to increase their visibility (see \cite{matosin} and \cite{franco}).
An important result that emerged from this study is the higher publication rate of programmes associated with larger allocations of telescope time. This is detected in both observing modes (SM and VM) as well as in the Normal programme type sub-sample. The SNPP did not reveal any significant dependence on allocated time in the distributions of responses for programmes with no refereed publications. This may be interpreted as an indication that a minimum amount of data is required to achieve results of a sufficient quality and quantity to warrant a publication (including the necessary effort that goes with it) across all science cases. We cannot exclude the possibility that the time distribution is skewed towards smaller requests by the general perception that this increases the chances of success rate during the selection process.
As the scientific process requires experimentation, it is necessary for an observatory to accommodate a fraction of risky proposals. When compounded with technical and weather losses, a 100 \% return in publications across all programmes becomes impossible. Nevertheless, the current level of 75 \% may be improved
by a further 10-15 \% by addressing specific factors. For example, by further optimising how observations are scheduled and executed at the telescope and re-evaluating the optimal fraction of risky observations, ESO can improve its data delivery performance. At the same time, the community can optimise the distribution of resources to ensure that data can be analysed more effectively as soon as it becomes available.

\acknowledgements

The authors are grateful to Francesca Primas, Martino Romaniello, and all of the members of the Time Allocation Working Group for their help during the formulation of the SNPP questionnaire.

\appendix

\section{Publication fraction time dependence}
\label{sec:app}

Let us first consider a single generation of programmes, all allocated at the same time. Let then $N_T$ be the total number of programmes that can produce a publication, $N_P$ the number of programmes that will eventually produce a publication, and $f^0_P=N_P/N_T$ the average publication fraction. With these settings, the number of programmes that will never produce a publication is $N_N=N_T \; (1-f^0_P)$. Let us then define $C(t)$ as the cumulative distribution function of the publication delay time distribution (PDTD):

\begin{displaymath}
C(t)= \int_0^t PDTD(x) \; dx
\end{displaymath}

The number of programmes that have already produced a publication at time $t$ is then:

\begin{equation}
\label{eq:np} 
N_P(t)= N_T f^0_P C(t)
\end{equation}

so that the number of programmes that are still working at time $t$ is:

\begin{displaymath}
N_W(t)= N_T \; f^0_P [1-C(t)]
\end{displaymath}

The number of programmes that have not published at time $t$ is $N_{NP}(t)=N_N + N_W(t)$, which can be written as:

\begin{displaymath}
N_{NP}(t) = N_T [1-f^0_p C(t)]
\end{displaymath}

We can now define the ratio between $N_W(t)$ and $N_{N}(t)$:

\begin{equation}
\label{eq:rt}
R(t) = f^0_P \frac{1-C(t)}{1-f^0_P C(t)}
\end{equation}

which is a quantity that can be directly measured.

If we now consider a set of multiple programme generations, the number of programmes still working on the data at time $t$ is given by:

\begin{displaymath}
N_W(t) = f^0_P \;\sum_i N_{T,i} [1-C(t-t_i)]
\end{displaymath}

where $N_{T,i}$ is the total number of programmes in the $i$-th generation that can produce a publication, and $t_i$ is the time when this generation was allocated. If we pose
$S=\sum_i N_{T,i}$ and $S_W=\sum_i N_{T,i}\;C(t-t_i)$, then we can write:

\begin{displaymath}
R(t) = f^0_P \frac{S-S_W}{S-f^0_P S_W}
\end{displaymath}

or more concisely as:

\begin{equation}
\label{eq:r}
R(t) = f^0_P \frac{1-\bar{C}}{1-f^0_P \bar{C}}
\end{equation}

where:

\begin{displaymath}
\bar{C} = \frac{S_W}{S} = \frac{\sum_i N_{T,i} C(t-t_i)}{\sum_i{N_{T,i}}}
\end{displaymath}

is the weighted mean of $C(t)$ over the duration of the survey, in which the weights are the number of proposals that can produce a publication in the given period. Therefore,
the expression for publication fraction for a population including different project generations is analogous to that of a single generation (Equation~\ref{eq:rt}), in which $C(t)$ is replaced
by its average weighted over the time range between the first generation and the time of the survey.

Equation~\ref{eq:r} can be readily inverted to yield:

\begin{equation}
f^0_P = \frac{R(t)}{1- \bar{C}[1-R(t)] }
\end{equation}

With similar considerations, one can generalise Equation~\ref{eq:np} for the whole multi-generation set of programmes:

\begin{displaymath}
\bar{N}_P (t)= N f^0_P \bar{C}
\end{displaymath}

where $N=\sum_i N_{T,i}$ is the total number of programmes that can produce a publication. Considering that $\bar{N}_P(t)/N=f_P(t)$, this finally yields:

\begin{displaymath}
f_P(t) = f^0_P \; \bar{C}
\end{displaymath}

which gives the expected publication fraction at the time of the survey.

\end{document}